\documentclass{PoS}

\font\twelvebb=msbm12
\font\tenbb=msbm10
\font\sevenbb=msbm7
  \newfam\bbfam
  \def\bb{\fam\bbfam\twelvebb}
  \textfont\bbfam=\twelvebb
  \scriptfont\bbfam=\tenbb
  \scriptscriptfont\bbfam=\sevenbb
\font\twelveeusm=eusm10 scaled 1200
\font\teneusm=eusm10
  \newfam\eusmfam
  
  \textfont\eusmfam=\twelveeusm
  \scriptfont\eusmfam=\teneusm
  \scriptscriptfont\eusmfam=\scriptfont\eusmfam
\font\twelvefrak=eufm10 scaled 1200
\font\tenfrak=eufm10
  \newfam\frakfam
  
  \textfont\frakfam=\twelvefrak
  \scriptfont\frakfam=\tenfrak
  \scriptscriptfont\frakfam=\scriptfont\frakfam
\def\sqr#1#2{{\vcenter{\hrule height.#2pt
   \hbox{\vrule width.#2pt height#1pt \kern#1pt
      \vrule width.#2pt}
   \hrule height.#2pt}}}

\def\bsqr#1#2{{\vrule width #1pt height#2pt}}
\def\bsquare{{\mathchoice\bsqr66\bsqr66\bsqr33\bsqr33}}
\def\badbreak{\penalty1000}

\def\Trs{\mathop{\rm tr}}                   
\def\Trb{\mathop{\rm Tr}}                   
\def\Det{\mathop{\rm det}}                  
\def\identity{{\bb I}}                      
\def\sgn{\mathop{\rm sgn}}                  
\def\rational#1#2{{\mathchoice{\textstyle{#1\over#2}}%
  {\scriptstyle{#1\over#2}}{\scriptscriptstyle{#1\over#2}}{#1/#2}}}
\def\third{\rational13}                     
\def\R{{\bb R}}                             
\newcommand{\gfive}{\gamma_{5}}             
\newcommand{\cO}{{\cal O}}                  
\newcommand{\psibar}{{\bar\psi}}            
\newcommand{\mbar}{{\bar m}}                
\newcommand{\Ubar}{{\bar U}}                
\newcommand{\barbeta}{{\bar \beta}}         
\newcommand{\bartheta}{{\bar \theta}}       

\title{Coherent lattice QCD}

\ShortTitle{Coherent lattice QCD}

\author{\speaker{Ivan Horv\'ath}\\
        Department of Physics and Astronomy,
        University of Kentucky,
        Lexington, KY 40503\\
        E-mail: \email{horvath@pa.uky.edu}}

\abstract{We discuss a proposal for the construction of lattice QCD 
  with gauge action, fermionic action, $\theta$--term, and the operators all 
  based on the lattice Dirac operator $D$ with exact chiral symmetry. The 
  simplest regularization of this type uses the proposition that the classical 
  limit of scalar gauge density associated with trace of $D$ is (up to an 
  additive constant) proportional to $\Trs F^2$, while the corresponding 
  operator is local. More general formulations from this class are considered 
  with the aim of exposing interrelations between gauge and fermionic aspects 
  of QCD which are otherwise hidden in generic formulations. Possible utility  
  of these formulations for exploring QCD vacuum structure is emphasized.}

\FullConference{XXIVth International Symposium on Lattice Field Theory\\
                July 23-28, 2006\\
                Tucson, Arizona, USA}

\begin{document}

\section{Vacuum Structure Motivation}

When studying QCD vacuum structure in the path integral formalism, the 
basic issue we face is the apparent clash of two tendencies. On one side there is 
{\em randomness} that is inherent in the definition of the theory, while on the 
other side there is {\em space--time order} that we seek in order to understand 
the vacuum. The root of the problem is largely related to the fact that we cannot 
completely eliminate one or the other. Indeed, eliminating the randomness seems
to require abandoning the field theory description of strong interactions
-- a step that is difficult to contemplate otherwise. On the other hand, 
the absence of the need for the element of space--time order would imply that 
vacuum cannot be understood in the language of path integrals. This means that
the analysis of some equilibrium configuration $U$ in lattice--regularized theory 
should not only be able to distinguish randomness from order, but it should also  
separate the randomness that is physically necessary from one that is useless, 
and can thus be eliminated. This is a highly non--trivial problem.

The proper formulation of the tasks qualitatively described above requires the use 
of concepts well established in the information theory and the theory of 
computation. A particular fusion of these subjects known as 
{\em algorithmic information theory} seems to be particularly 
relevant~\cite{Hor06A}. At the same time, 
once the underlying issues are properly identified, one can also attempt a 
``physicist's approach'' to the problem, trying to take advantage of possible 
shortcuts offered by the proper physical insight. Following this path within 
the {\em ``Bottom-Up''} approach to QCD vacuum structure~\cite{Hor03A,Hor06A}, 
we will rely on two guiding points. (1) Since truly random fluctuations cannot
be spatially correlated, it should be possible to remove them {\em locally}.
(2) Local transformations devised for such purposes should preserve the 
local physical meaning of the gauge field, thus giving a reasonable expectation
that only unphysical randomness is eliminated while physics is retained
(together with the randomness that is necessary). The first candidate 
transformations of this kind are the {\em chiral ordering transformations} 
of Ref.~\cite{Hor06A}. Given a chirally symmetric Dirac operator $D$, 
the prototypical chiral ordering transformation replaces the link $U_{n,\mu}$ 
with an SU(3) element $\Ubar_{n,\mu}$ minimizing the norm of the 
matrix~\cite{Hor06A}
\begin{equation}
     D_{n,n+\mu}(U) \,-\, D^f_\mu \times \Ubar_{n,\mu}  \qquad\quad
     \mbox{\rm where}  \qquad\quad
     D^f_\mu \equiv \third {\Trs}^c D_{n,n+\mu}(\identity) 
     \label{eq:2}
\end{equation}
In other words, the transformed gauge connection represents an effective
matrix phase by which fermionic variables get rotated under hopping from 
$n+\mu$ to $n$, relative to the free hopping. This preserves the local 
interpretation of the gauge field as a ``phase rotator'' for the charged 
particle interacting with it. Given the expected non--ultralocality of $D$ 
in gauge variables, the above transformations are non--trivial for generic 
backgrounds.

Using the locality of chiral ordering transformations, one can argue that 
applying them repeatedly at the ensemble level leads to the evolution in the 
set of valid gauge actions~\cite{Hor06A}. The theories reached this way are 
expected to generate more space--time order (lower Kolmogorov entropy) in their 
representative configurations than the starting theory. The characteristic 
feature of corresponding actions is that they become functions of the chirally 
symmetric Dirac kernel $D$ on which the transformation is based. This fact 
suggests a tempting possibility that, at least for the purposes of studying 
the QCD vacuum structure, it would be interesting to explicitly construct 
lattice regularizations where both gauge and fermionic parts of the theory 
enter in a unified manner, namely via lattice Dirac operator with exact chiral 
symmetry~\cite{Hor06B}. In fact, such {\em coherence} can be extended to all 
elements of the theory including the operators for relevant observables. 
In this talk we discuss some aspects of this construction, given fully in 
Ref.~\cite{Hor06B}, with the particular focus on arriving at the formulation
where gauge and fermionic parts of the full action are cast in the most 
mutually symmetric form.

\section{Simple Coherent Lattice QCD}

The simplest version of lattice QCD (LQCD) where all parts of the action are 
constructed from $D$ ({\em coherent LQCD}), is based on the following 
conjecture~\cite{Hor06B}~\footnote{Note that we will keep both the notation 
and the numbering of conjectures the same as in Refs.~\cite{Hor06A,Hor06B}.
Also, the local traces are denoted by ``$\Trs$'' while the global ones by ``$\Trb$'' 
and, unless denoted explicitly by a superscript, the local traces are taken 
over the full linear space appropriate for the object in question.}
\medskip

\noindent {\bf Conjecture C3.} {\em Let $A_\mu(x)$ be arbitrary smooth  {\rm su(3)} 
gauge potentials on $\R^4$. If $U(a) \equiv \{\,U_{n,\mu}(a)\,\}$ is the transcription 
of this 
field to the hypercubic lattice with classical lattice spacing $a$, and 
$\identity \equiv \{\,U_{n,\mu} \rightarrow \identity^c\,\}$ is the free field 
configuration then 
\begin{equation}
   \Trs\, \Bigl( D_{0,0}(U(a)) \,-\, D_{0,0}(\identity) \Bigr) \;=\;
   -c^S a^4\, \Trs F_{\mu\nu}(0) F_{\mu\nu}(0) \,+\, \cO(a^6)
   \label{eq:5}
\end{equation} 
for generic chirally symmetric $D$. Here $c^S$ is a non--zero constant independent of 
$A_\mu(x)$ at fixed $D$, and 
$F_{\mu\nu}(x) \,\equiv\, 
      \partial_\mu A_\nu(x) - \partial_\nu A_\mu(x) + [\, A_\mu(x), A_\nu(x) \,]$ is 
the field--strength tensor.}
\medskip

\noindent
The heuristic reason for validity of the above conjecture is that $\Trs D_{n,n}$ is
scalar, local, gauge invariant function of the gauge field. Up to dimension four, 
the only possibilities for such operators in the continuum are the constant and
$\Trs F_{\mu\nu}(x)F_{\mu\nu}(x)$. 
For the family of standard overlap Dirac operators based on the Wilson-Dirac operator 
with mass $-\rho$~\cite{Neu98BA}, the validity of Conjecture C3 will be supported 
by both analytical and numerical methods in Ref.~\cite{Ale06A}, where the constants 
$c^S(\rho)$ will also be evaluated. Note that the conjecture analogous to C3 in 
the pseudoscalar case is in fact a basis for constructing 
topological density from chirally symmetric Dirac operator~\cite{Has98A} 
(see also Ref.~\cite{NarNeu95}). For overlap Dirac operator this was examined
explicitly in Refs.~\cite{c_pscalar}.

Accepting the validity of Conjecture C3 allows us to define the action for simplest 
version of coherent LQCD with $N_f$ flavors of quarks and the CP--violating 
$\theta$--term as~\cite{Hor06B}
\begin{equation}
   S_{\barbeta,\bartheta,\{m_f\}}(U) \,\;=\;\,
   \Trb\, ( \barbeta - i \bartheta \gfive) \Bigl( D(U) - D(\identity) \Bigr) \;+\;
   \sum_{f=1}^{N_f} \psibar^f \Bigl( D(U) + m_f \Bigr) \psi^f
   \label{eq:30} 
\end{equation}
where $\{m_f\}$ is the set of real non--negative quark masses, and
\begin{equation}
   \barbeta(\beta)  \,=\, \frac{\beta}{12\, c^S}   \qquad
   \Bigl( \,\beta \equiv \frac{6}{g^2} \, \Bigr)  \qquad\qquad\;
   \bartheta(\theta) \,=\, \frac{\theta}{16 \pi^2 c^P}   \qquad
   \Bigl( \, \theta \in (-\pi,\pi] \, \Bigr)   
   \label{eq:35}
\end{equation}
The constant $c^P$ appearing in pseudoscalar term is defined analogously to $c^S$. 
We note that the locality of $S$ defined above follows from locality of $D$, 
and that the free--field term in the gauge part of the action only contributes 
a field--independent constant which can be discarded if desired.

The dynamics of coherent QCD (\ref{eq:30}) is completely encoded in the chirally 
symmetric lattice Dirac operator. This becomes more explicit after fermionic 
variables are integrated out. In case of degenerate quark masses the distribution 
density of the gauge fields is given by
\begin{equation}
    P_{\barbeta,\bartheta,m}(U)  \; \propto \;
    e^{\Trb [\, N_f \ln ( D + m ) \,+\, 
               (-\barbeta + i \bartheta \gfive) D \,]} \;=\;
       \Det\,\Bigl[\, (D + m)^{N_f}\, 
       e^{(-\barbeta + i \bartheta \gfive) D } \,\Bigr] 
    \label{eq:40}
\end{equation}
where $D\equiv D(U)$. Thus, for $\theta=0$ this formulation requires simulating
the probability distribution
\begin{equation}
    P_{\barbeta,m}(U)  \; \propto \;
       \Det\,\Bigl[\, (D + m)^{N_f}\, 
       e^{-\barbeta D } \,\Bigr]  \;=\;
       \Det\,\Bigl[\,\Bigl(-\frac{d}{d\barbeta}\Bigr)^{N_f}\, 
       e^{-\barbeta (D+m)} \,\Bigr] 
       \label{eq:41}
\end{equation}
where the last equality emphasizes the close explicit relation between fermionic
and gauge contributions to the full probability distribution. Some initial ideas
on simulating this theory based on overlap Dirac kernel are presented in
Ref.~\cite{KFL06B}.

\section{Symmetric Logarithmic LQCD}

It is easy to see that the simple coherent LQCD discussed above is by no means
the only possibility for constructing lattice regularizations where the gauge 
and fermionic parts of the action are tied together in an explicit manner. 
To obtain more general class of such formulations, we will treat the Dirac 
operator $D$ defining fermionic action as a primary object, and discuss 
possibilities for forming the gauge parts of the action from various functions 
$f(D)$. In particular, if $f(D)$ is a local operator and if $\Trs f(D)_{n,n}$ 
is a scalar lattice field, then we generically expect the validity of 
a statement analogous to Conjecture C3 with equation (\ref{eq:5}) replaced by
\begin{equation}
   \Trs\, \Bigl[ f\Bigl( D(U(a)) \Bigr)_{0,0} \,-\, 
                 f\Bigl( D(\identity) \Bigr)_{0,0} \,\Bigr] \;=\;
   -c^S a^4\, \Trs F_{\mu\nu}(0) F_{\mu\nu}(0) \,+\, \cO(a^6)
   \label{eq:55}
\end{equation} 
where $c^S$ is an associated constant. Proceeding in the same way also for 
the pseudoscalar case (i.e. considering operator $\gfive f(D)$) we obtain the 
definition of coherent LQCD in this case by replacing $D$ in the gauge part 
of Eq.~(\ref{eq:30}) with $f(D)$, while Eq.~(\ref{eq:35}) remains unchanged.
Some attractive choices for functions $f(D)$ include
$D^k,\; \ln(D+\eta),$ and $(D+\eta)^{-1}$~\cite{Hor06B}.

We will now focus on constructing coherent LQCD in which the gauge and fermionic
contributions to the action enter in the most mutually symmetric manner. Since 
the fermionic part is fixed, we can bring the two forms closer together if we
mimic the effective fermionic action (after integrating out fermions) in the
construction of the gauge part, i.e. if we use $f(D)=\ln(D+\eta)$. This leads to 
\begin{equation}
   S_{\barbeta,\bartheta,\{m_f\}}(U) \;=\; 
   \Trb\, ( \barbeta - i \bartheta \gfive) \ln \Bigl( D(U) + m_0 \Bigr) 
   \;+\;
   \sum_{f=1}^{N_f} \psibar^f \Bigl( D(U) + m_f \Bigr) \psi^f
   \label{eq:60} 
\end{equation}
where we denoted $\eta\equiv m_0>0$ to reflect its mass--like form. Note
however that, unlike the lattice quark masses $m_f$, the parameter $m_0$ is kept 
fixed as the continuum limit is approached. In effect, different values of 
$m_0$ control the lattice locality range of the gauge action. We refer to the above
formulation as {\em logarithmic} LQCD~\cite{Hor06B}. For $\theta=0$ and 
degenerate quark masses, the probability distribution of gauge fields 
to simulate in this case is
\begin{equation}
    P_{\barbeta,m}(U)  \; \propto \;
    e^{\Trb [\, N_f \ln ( D + m ) \,-\, 
    \barbeta \ln ( D+m_0 ) \,]}  \;=\;
    \Det\,\Bigl[\, (D + m)^{N_f}\, (D+m_0)^{-\barbeta} \,\Bigr]
    \label{eq:62}
\end{equation}

Interestingly, it is possible to bring the gauge action to the form that is yet 
more (and completely) analogous to the effective action of a single lattice
fermionic flavor. To see that, consider the effective action of logarithmic LQCD 
with one flavor and $\theta=0$, namely
\begin{equation}
    -S^{eff}_{m,\barbeta}(U) \;=\; \Trb \,[\,\ln (D+m) -
                                   \barbeta \,\ln (D+m_0)\,]
    \qquad\qquad
    \barbeta \equiv \frac{1}{2g^2 c^S(m_0)} 
    \label{eq:63}
\end{equation}
We can thus eliminate varying $\barbeta$ in favor of varying
$m_0$ (and bring the two terms into almost identical form), if for arbitrary $g>0$ 
from some finite vicinity of $g=0$, we can find unique $m_0\equiv m_0(g)$ such that 
$\, |c^S(m_0)| \equiv 1/2g^2$. This is indeed expected to be possible since 
$\ln (D+m_0)_{n,n}$ will diverge as $m_0\rightarrow 0$ 
for smooth configurations. A precise statement can be formulated as 
follows~\cite{Hor06B}.
\smallskip

\noindent {\bf Conjecture C6.} {\em Let $D$ be a chirally symmetric operator 
such that $f(D) \equiv \ln (D+\eta)$ is well--defined for arbitrary $\eta>0$. 
If $c^S(\eta)$ is the associated classical coupling of $\Trs\!f(D(U))_{n,n}$ 
to $\Trs F_{\mu\nu}F_{\mu\nu}$, then there exists $\eta_0>0$ such that 
$c^S(\eta)$ is monotonic for $0<\eta\le \eta_0$, and
$\lim_{\eta \to 0}\, |c^S(\eta)| \,=\,\infty$. Moreover, there exists
a non--zero (possibly infinite) limit
\begin{equation}
    \lim_{\eta \to 0} \; \frac{c^S(\eta)}{\ln (\eta)} 
    \,\equiv\, \lim_{\eta \to 0} \kappa^S(\eta) 
    \,\equiv\, \kappa^S(0) \ne 0
    \label{eq:64}
\end{equation}
}
Conjecture C6 implies the existence of a one-to-one correspondence between 
$g \in (0,g_0]$ and $m_0 \in (0,\eta_0]$, such that 
$\, |c^S(m_0)| \equiv 1/2g^2$. We thus have instead of (\ref{eq:63})
\begin{equation}
    -S^{eff}_{m,m_0}(U) \,=\, \Trb \,[\,\ln (D+m) \,-\, 
                               \sgn(c^S(m_0)) \,\ln (D+m_0)\,]
    \qquad\quad
    |c^S(m_0)| \equiv \frac{1}{2g^2}      
    \label{eq:66}
\end{equation}

The full--fledged form of the lattice theory constructed according to the 
above arguments is~\cite{Hor06B}
 \begin{eqnarray}
   S_{m_0^S,m_0^P,\{m_f\}} &=& 
     \,\sgn\Bigl( c^S(m_0^S) \Bigr) \Trb\, \ln \Bigl( D + m_0^S \Bigr) 
     \,-\, 
     i \,\sgn\Bigl( c^P(|m_0^P|) \Bigr) \,\sgn(m_0^P) \Trb\, 
       \gfive \ln \Bigl( D + |m_0^P| \Bigr)
       \nonumber \\
   &+& \sum_{f=1}^{N_f} \psibar^f \Bigl( D + m_f \Bigr) \psi^f
   \label{eq:105} 
\end{eqnarray}
where the ``gauge'' parameters $m_0^S$, $m_0^P$ are related to $g$ and 
$\theta$ via
\begin{equation}
    |c^S(m_0^S)| \equiv \frac{1}{2g^2}      \qquad\qquad
    \theta \,=\, \sgn(m_0^P)\, 16 \pi^2 \,|c^P(|m_0^P|)|  \qquad\qquad 
    \theta \in (-\pi,\pi]
    \label{eq:106}
\end{equation}
and the various mass--like lattice parameters of the theory vary within 
the ranges
\begin{equation}
     m_f \in (0,\infty)    \qquad\qquad
     m_0^S \in (0,m_0^{S,c}\,] \qquad\qquad
     m_0^P \in [\,m_0^{P,c},\infty) \cup (-\infty,-m_0^{P,c}\,)
     \label{eq:107}
\end{equation}
In the above equations, $m_0^{S,c}$ is the maximal $\mu_0$ satisfying the statement of 
Conjecture C6. Note that in this formulation we have allowed the pseudoscalar mass 
$m_0^P$ to be negative so that the lattice action density preserves exactly 
the transformation property of the pseudoscalar part under 
$\theta \rightarrow -\,\theta$. Such definition assumes that $|c^P(\eta)|$ vanishes 
at infinity and is monotonically decreasing in the range $\eta\in [m_0^{P,c},\infty)$, 
where $m_0^{P,c}$ is defined via 
$\theta(m_0^{P,c})\equiv \pi = 16 \pi^2 \,|c^P(m_0^{P,c})|$. For the family 
of overlap Dirac operators $D^{(\rho)}$ this is satisfied with
$m_0^{P,c} \;=\; \frac{2\rho}{e^\pi -1}$. Lattice regularization constructed 
above is referred to as {\em symmetric logarithmic} LQCD~\cite{Hor06B}. 
The continuum limit with $N_f \le 16$ is taken via 
$m_0^S \equiv m_0^S(a) \rightarrow 0$ while decreasing $m_f=m_f(a) \propto a\mbar_f^r$ 
towards zero so that some set of renormalized masses $\mbar_f^r$ (in physical units) is 
held fixed, and with the pseudoscalar mass $m_0^P$ (specifying $\theta$) kept unchanged
in the process.

From Eq.~(\ref{eq:66}) one can see that the gauge action in symmetric logarithmic LQCD
can be viewed as an effective action of a lattice fermion or pseudofermion, depending
on the sign of $c^S(\eta)$ in the vicinity of $\eta=0$. Preliminary calculations 
indicate~\cite{Ale06A} that, for overlap Dirac operator, $c^S(\eta)<0$ sufficiently 
close to $\eta=0$, and this is expected to be true in general. In this case,
the effective action of symmetric logarithmic LQCD at $\theta=0$ can be written as 
\begin{equation}
   -S^{eff}_{\{m_f\}} \,=\, 
    \Trb \, \sum_{f=0}^{N_f} \ln\, \Bigl( D(U) \,+\,m_f \Bigr)
    \,=\,
    \Trb \, \ln \prod_{f=0}^{N_f} \Bigl( D(U) \,+\,m_f \Bigr)
    \label{eq:109}
\end{equation}
with the gauge contribution to the action entering at the regularized level
in a completely form--symmetric manner relative to the contribution of a single 
fermionic flavor. Introducing the Grassmann variables for this ``0--th flavor''
of mass $m_0\equiv m_0^S$, the total action can be written in the form
\begin{equation}
   S_{\{m_f\}} \;=\; 
   \sum_{f=0}^{N_f}\, \psibar^f \Bigl( D(U) + m_f \Bigr) \psi^f
   \label{eq:190}
\end{equation}
Thus, symmetric logarithmic LQCD casts the regularized dynamics of full 
QCD into that of $N_f+1$ lattice fermionic flavors interacting with SU(3) gauge 
field. While this might sound suspicious at first, in fact it is not. It turns 
out that what distinguishes ``gauge'' from ``fermionic'' in this case is the 
locality of corresponding effective operators in the continuum limit. 
Indeed, since for the ``0--th'' flavor the corresponding effective operator is 
the gauge action itself, we have to insist that its effective range in physical 
units shrinks to zero in the continuum limit (``weak locality''~\cite{Hor06B}). 
For theory with $N_f$ asymptotically free flavors this translates into 
the requirement~\cite{Hor06B}
\begin{equation}
   0 \;=\; \lim_{a \to 0} \frac{a}{m_0(a)} \;\propto\; 
           \lim_{m_0 \to 0} 
           \;m_0^{\frac{|\kappa^S(0)|}{\beta_0}-1}
           \qquad \mbox{\rm or} \qquad
           \frac{|\kappa^S(0)|}{\beta_0} >1
   \label{eq:191}       
\end{equation}
where $\beta_0 = (11-\frac{2}{3}N_f)/16\pi^2$. The above condition can also
be viewed as a requirement on the number of asymptotically free flavors for 
which QCD can be defined via symmetric logarithmic LQCD, namely that
$N_f >  \frac{33}{2} \,-\, 24\,\pi^2\, |\kappa^S(0)|$. Thus, if this 
condition is satisfied, then the range of the effective action for
the ``0--th'' (gauge) flavor is zero in the continuum limit, while 
the ranges of effective actions corresponding to usual quark flavors are
non--zero and inversely proportional to the corresponding renormalized quark 
masses $\mbar_f^r$. This is equivalent to saying that the ``0-th'' flavor 
in (\ref{eq:190}) is infinitely heavy in the continuum limit. Indeed, from
Eq.~(\ref{eq:191}) we have that its mass in physical units is 
proportional to
\begin{equation}
     \lim_{a \to 0} \frac{m_0(a)}{a} \,\propto\, 
     \lim_{m_0 \to 0} m_0^{1-\frac{|\kappa^S(0)|}{\beta_0}} \,=\,
     \infty
\end{equation}
Consequently, all correlators involving the variables $\psi_0$, $\psibar_0$
in theory (\ref{eq:190}) are expected to vanish in the continuum limit, and 
the ``0--th'' flavor decouples from the light quark flavors.

\section{Discussion}

In this talk we have argued that, for the purposes of studying QCD vacuum 
structure, it might be fruitful to explore lattice regularizations of QCD
where gauge and fermionic aspects of the theory are both based on a lattice
Dirac operator $D$ with exact chiral symmetry (coherent LQCD)~\cite{Hor06B}.
The motivations associated with vacuum structure are mostly related to 
the relevance of chiral ordering transformations of Ref.~\cite{Hor06A}.  
However, it is quite clear that this novel approach to constructing 
lattice regularizations might be interesting in its own right. Indeed,
the explicit interrelations between gauge and fermionic aspects of the theory,
present in such regularizations, potentially offer valuable insights into
the ``inner workings'' of QCD dynamics. After all, the properties of 
valid regularizations become properties of QCD in the continuum limit. 
Following this route, the goal of our discussion was to arrive at 
the formulation where gauge and fermionic parts of the full action become 
mutually form--symmetric to a maximal possible degree. 
Starting from simplest coherent LQCD and its generalizations, we 
arrived at symmetric logarithmic LQCD, where quarks and gluons contribute
to an overall dynamics in a completely form--symmetric manner~\cite{Hor06B}. 
This can be most clearly seen in Eq.~(\ref{eq:190}), where the gauge part
of the dynamics is represented via an additional fermionic flavor that becomes
infinitely heavy in the continuum limit.

The existence of lattice regularizations where gauge dynamics can be viewed 
as inherited from infinitely heavy fermions (in logarithmic LQCD there  
are in fact infinitely many of them in the continuum limit~\cite{Hor06B})
raises a question if QCD has a ``natural definition'' from the point of view
of the theory beyond the Standard Model. Indeed, if the heavy decoupled 
particles in question can be given physical meaning in such context, then
this possibility might acquire some content.

Finally, we wish to mention that the ``coherence'' in definition of LQCD 
can be extended to include all operators by application of chiral ordering
transformations~\cite{Hor06A,Hor06B}. Some interesting operators can also 
be constructed from $D$ explicitly~\cite{Hor06B,KFL06B,Ale06A}. 

\smallskip

\noindent {\bf Acknowledgment:} Numerous discussions with Andrei Alexandru 
and Keh-Fei Liu on the topics related to the subject of this talk are 
gratefully acknowledged.

\end{document}